\documentclass[12pt]{article}
\usepackage{amsmath}
\usepackage{graphicx}
\usepackage{a4}
\usepackage{amsfonts}
\usepackage{amssymb}

\csname @addtoreset\endcsname{equation}{section}
\newcommand{\eq}{\begin{equation}}
\newcommand{\feq}{\end{equation}}
\newcommand{\eqn}{\begin{eqnarray}}
\newcommand{\feqn}{\end{eqnarray}}
\newcommand{\arr}{\begin{eqnarray*}}
\newcommand{\farr}{\end{eqnarray*}}

\begin{document}
\begin{titlepage}
\begin{flushright}
CAMS/01-05\\
\end{flushright}
\vspace{.3cm}
\begin{center}
\renewcommand{\thefootnote}{\fnsymbol{footnote}}
{\Large\bf
Einstein Brane-Worlds In 5D  Gauged Supergravity}
\vskip20mm
{\large\bf{A. H. Chamseddine \footnote{email: chams@aub.edu.lb} and
W.~A.~Sabra\footnote{email: ws00@aub.edu.lb}}}\\
\renewcommand{\thefootnote}{\arabic{footnote}}
\vskip2cm
{\it
Center for Advanced Mathematical Sciences (CAMS)
and\\
Physics Department, American University of Beirut, Lebanon.\\}
\end{center}
\vfill
\begin{center}
{\bf Abstract}
\end{center}
\vskip1cm
We study, in the context of five dimensional $N=2$ gauged supergravity with vector
and hypermultiplets, curved domain wall solutions with worldvolumes given by four
dimensional Einstein manifolds. For a choice of the projection condition
on the Killing spinors of the BPS solutions, first order differential equations
governing the flow of the scalars are derived. With these equations, we analyze
the equations of motion and
determine conditions under which gauged supergravity theories may
admit Einstein domain wall solutions.
\end{titlepage}

\section{Introduction}

Recently there has been some interest in the study of supersymmetric as well
as non-supersymmetric domain walls and black holes as solutions of $N=2$
five-dimensional gauged supergravity. This activity has been motivated mainly
by the desire to embed the Randall-Sundrum \cite{RS} scenario in the framework
of string or M-theory. The choice for five dimensional gauged theories comes
from the fact that such theories allow for anti-de Sitter vacuum states which
are fundamental for the realization of the Randall-Sundrum model. Also of
interest is the study of black hole solutions in gauged supergravity theories
as they play a fundamental role in the conjectured AdS/CFT correspondence
\cite{ads}. In the past few years, \ supersymmetric black holes and strings,
as well as non-supersymmetric generalizations, have been constructed (see
\cite{bcs2}) for the $U(1)$-gauged $N=2$ supergravity \cite{gst2}.

Very recently, there has been a shift towards the study of $N=2$ supergravity
with gauged hypermultiplets. Ultimately, one would like to investigate the
most general theories and study their solutions in the hope of finding a
particular model which may incorporate Randall-Sundrum scenario in a
supersymmetric setting. Flat domain walls and black hole solutions for five
dimensional supergravity theories with gauged isometries of the
hypermultiplets have been discussed very recently in \cite{domaink, BC,
gutsabra}.

In this paper we are interested in the study of curved Einstein domain walls
for the gauged supergravity theories discussed in \cite{ceresole}. Such a
study has been initiated by the recent work of Cardoso et al \cite{lust}. In a
previous paper \cite{chamsabra}, we have shown that the flat-worldvolume
domain wall solutions found in \cite{domaink} can be generalized to solutions
with Ricci-flat worldvolumes. In the work of \cite{domaink} it was established
that flat BPS domain wall solutions of gauged supergravity with
hypermultiplets can only exist under certain conditions (see next section).
These conditions are in general not satisfied and this may restrict the class
of gauged supergravities that have flat BPS solutions. Our purpose in this
work is the study of supersymmetric domain wall solutions with four
dimensional worldvolumes given by Einstein spaces with a negative cosmological
constant. In the analysis of \cite{chamsabra}, where the projection condition
on the Killing spinors as given in \cite{domaink} was used, it was shown that
such solutions do not exist. Later, and in the revised work of \cite{lust} a
more general projection condition was proposed. We shall show here that
supersymmetric Einstein domain wall solutions in presence of non-trivial
matter, within the framework of \cite{lust}, may be allowed under stringent
conditions. However, non-supersymmetric solutions may be possible if one
satisfies the equations of motion and ignores the integrability conditions
coming from the Killing spinors equations. The conditions under which the
non-supersymmetric solutions of \cite{karch} (generalized to many vector and
hypermultiplets) can be obtained as solutions of five dimensional supergravity
theory are determined. Our analysis is carried out in the context of gauged
$N=2$ supergravity models of \cite{ceresole} and in the absence of tensormultiplets.

We organize this work as follows. In the next section the supergravity
theories we wish to study are reviewed together with a brief discussion on
their possible Ricci-flat domain wall solutions. In section three, we look for
BPS Einstein domain wall solutions and derive first order differential
equations by solving for the vanishing of supersymmetry transformations of the
fermi fields in a bosonic background. This is done for a general choice of the
projection condition on the Killing spinors which was given in \cite{lust}.
The equations of motion are analyzed and we determine some conditions under
which the models presented in \cite{karch} can be embedded in gauged $N=2$
supergravity theory. We demonstrate that the constraints derived from the
equations of motion when combined with the integrability conditions (coming
from the requirement of unbroken supersymmetry) give strong constraints.
Finally we summarize our results and discuss possible future directions.

\section{Ricci-Flat Domain Walls}

In this section we briefly review gauged supergravity models and their
possible Ricci-flat domain wall solutions \cite{ceresole, domaink, chamsabra}.
The gauged supergravity theories we are interested in are those minimal
theories (with eight real supercharges) coupled to $n_{V\text{ }}%
$vectormultiplets and $n_{H}$ hypermultiplets, where global isometries
including $R$-symmetry are made local. Specifically, we consider the models
constructed in \cite{ceresole} without tensormultiplets. The fermionic fields
of the $N=2$ supergravity theory are \footnotetext[1]{In this paper, the
indices $A,B$ represent five-dimensional flat indices, $A=(a,5).$ Curved
indices are represented by $M$ =$(\mu$, $z).$} the gravitini $\psi_{M}^{i}$
which are symplectic Majorana spinors ($i=1,2$ are $SU_{R}(2)$ indices), the
gaugini $\lambda_{i}^{\hat{a}}$ \footnote[2]{Here $\hat{a}$ is the flat index
of the tangent space group $SO(n_{V})$ of the scalar manifold $\mathcal{M}%
_{V}.$} and the hyperini $\zeta^{\alpha}$ ($\alpha=1,...,2n_{H})$. The bosonic
fields consist of the graviton, vector bosons $A_{M}^{I}$ $(I=0,1,....,n_{V}%
)$, the real scalar fields $\phi^{x}$, ($x=1,\ldots,n_{V})$ of the
vectormultiplets and the scalars $q^{X}$ ($X=1,...,4n_{H})$ of the
hypermultiplets. The scalar fields of the theory live on a manifold
$\mathcal{M}=\mathcal{M}_{V}\otimes\mathcal{M}_{H},$ which is the direct
product of a very special \cite{dWvP} and a quaternionic K\"{a}hler manifold
\cite{BW} with metrics denoted respectively by $g_{xy}(\phi)$ and $g_{XY}(q)$.
The target manifold of the scalar fields of the vectormultiplets
$\mathcal{M}_{V}$ is a very special manifold described by an $n_{V}%
$--dimensional cubic hypersurface
\begin{equation}
C_{IJK}h^{I}(\phi^{x})h^{J}(\phi^{x})h^{K}(\phi^{x})=1
\end{equation}
of an ambient space parametrized by $n_{V}+1$ coordinates $h^{I}=h^{I}%
(\phi^{x})$, where $C_{IJK}$ is a completely symmetric constant tensor which
defines Chern--Simons couplings of the vector fields. A classification of the
allowed homogeneous manifolds can be found in \cite{dWvP}. The quaternionic
K\"{a}hler manifold can be described in terms of the $4n_{H}$-beins
$f_{i\alpha}^{X}$ obeying the relation $g_{XY}\,f_{i\alpha}^{X}\,f_{j\beta
}^{Y}=\epsilon_{ij}\,C_{\alpha\beta}$, where $\epsilon_{ij}$ and
$C_{\alpha\beta}$ are the $SU(2)$ and $USp(2n_{H})$ invariant tensors respectively.

We are mainly interested in finding bosonic configurations and we display only
the bosonic action of the gauged theory as well as the supersymmetry
variations of the fermi fields in a bosonic background. The bosonic action for
vanishing gauge fields is given by
\begin{equation}
E^{-1}\mathcal{L}=\frac{1}{2}R-\frac{1}{2}g_{XY}\partial_{M}q^{X}\partial
^{M}q^{Y}-\frac{1}{2}g_{xy}\mathcal{\partial}_{M}\phi^{x}\mathcal{\partial
}^{M}\phi^{y}-\mathcal{V}(\phi,q),\label{smm}%
\end{equation}
where $E=\sqrt{-\det g_{MN}}$ \ and the scalar potential is given by
\cite{ceresole}
\[
\mathcal{V}=-g^{2}\left[  2P_{ij}P^{ij}-P_{ij}^{\hat{a}}P^{\hat{a}%
\,ij}\right]  +2g^{2}\mathcal{N}_{i\alpha}\mathcal{N}^{i\alpha}.
\]
and
\begin{equation}
P_{ij}\equiv h^{I}P_{I\,ij},\text{ \ }P_{ij}^{\hat{a}}\equiv h^{\hat{a}%
I}P_{I\,ij},\text{ \ \ \ }\mathcal{N}^{i\alpha}\equiv\frac{\sqrt{6}}{4}%
h^{I}K_{I}^{X}f_{X}^{\alpha i}.
\end{equation}
Here $K_{I}^{X},$ $P_{I\,}$ are the Killing vectors and prepotentials
respectively. For details of the gauging and the meaning of the various
quantities, we refer the reader to \cite{ceresole}.

In a bosonic background, the supersymmetry transformations of the fermi fields
in the gauged theory (after dropping the gauge fields contribution) are given
by
\begin{align}
\delta\psi_{Mi} &  =\mathcal{D}_{M}\varepsilon_{i}+\frac{i}{\sqrt{6}}%
g\,\Gamma_{M}\varepsilon^{j}P_{ij},\nonumber\\
\delta\lambda_{i}^{\hat{a}} &  =-\frac{i}{2}f_{x}^{\hat{a}}\Gamma
^{M}\varepsilon_{i}\,\partial_{M}\phi^{x}+g\varepsilon^{j}P_{ij}^{\hat{a}%
},\nonumber\\
\delta\zeta^{\alpha} &  =-\frac{i}{2}f_{iX}^{\alpha}\Gamma^{M}\varepsilon
^{i}\mathcal{\partial}_{M}q^{X}+g\varepsilon^{i}\mathcal{N}_{i}^{\alpha
}.\label{variation}%
\end{align}
Flat domain walls for the general gauged $N=2$ supergravity theory without
tensormultiplets were considered in \cite{domaink}. There, it was found that
if one writes
\begin{equation}
P^{(r)}(\phi,q)=h^{I}(\phi)P_{I}^{(r)}(q)=\sqrt{\frac{3}{2}}WQ^{(r)}%
,\ \text{\ \ \ \ }Q^{(r)}Q^{(r)}=1,
\end{equation}
where $W$ is the norm (the superpotential) and $Q^{\left(  r\right)  }$ are
$SU(2)$ phases of $P^{(r)}(\phi,q)$, then the existence of BPS flat domain
wall solutions with metric
\begin{equation}
ds^{2}=e^{2U(z)}\eta_{\mu\nu}dx^{\mu}dx^{\nu}+dz^{2},
\end{equation}
and Killing spinors satisfying
\begin{equation}
\Gamma_{z}\varepsilon_{i}=\gamma_{5}\varepsilon_{i}=Q^{(r)}\sigma_{ij}%
^{(r)}\varepsilon^{j},\label{proj}%
\end{equation}
will require that $Q^{\left(  r\right)  }$ satisfy the condition
\begin{equation}
\partial_{x}Q^{\left(  r\right)  }=0.\label{sidon}%
\end{equation}
The condition (\ref{sidon}) then implies that the scalar potential takes a
form which guarantees stability \cite{boucher},
\begin{equation}
\mathcal{V=}g^{2}(-6W^{2}+\frac{9}{2}g^{\Lambda\Sigma}\partial_{\Lambda
}W\partial_{\Sigma}W)\label{pottwo}%
\end{equation}
where $\Lambda,\Sigma$ run over all the scalars of the theory. The condition
(\ref{sidon}) is satisfied when there are no hyper scalars but only Abelian
vectormultiplets, in which case the $Q^{\left(  r\right)  }$ are constants.
Also (\ref{sidon}) is obviously satisfied when there are no physical
vectormultiplets. In general, the condition (\ref{sidon}) is not satisfied for
a generic point on the scalar manifold, and this may restrict the class of
gauged theories that have Ricci-flat BPS solutions.

The scalar fields and the warp factor for the flat BPS domain wall solutions
are given by \cite{domaink}
\begin{align}
\phi^{^{\prime}\Lambda}  &  =-3gg^{\Lambda\Sigma}\partial_{\Sigma}W,\text{
\ \ \ }\phi^{\Lambda}=(\phi^{x},q^{X}),\nonumber\\
U^{\prime}  &  =gW\text{.}\label{kalosh}%
\end{align}
The prime symbol denotes differentiation with respect to the fifth coordinate
$z$. As discussed in \cite{chamsabra}, the flat BPS domain walls of
\cite{domaink} can be promoted to solutions with Ricci-flat worldvolumes. The
amount of supersymmetry preserved by the five dimensional domain wall depends
on the amount of supersymmetry preserved by its four dimensional Ricci-flat worldvolume.

\section{Einstein Domain Walls}

In this section, we are interested in studying domain walls with Einstein
worldvolumes with a negative cosmological constant. In \cite{chamsabra}, it
was shown that supersymmetric solutions are not necessarily solutions of the
equations of motion (see also \cite{bls, aljose}). Therefore it is important,
in our subsequent analysis, to make sure that possible supersymmetric
configurations satisfy the equations of motion. Our main interest is to
determine the conditions under which the $N=2$ gauged supergravity theories
may allow for Einstein domain wall solutions. We start our analysis by
allowing for a general projection condition on the Killing spinors; therefore,
we write \cite{lust}
\begin{align}
\gamma_{5}\varepsilon_{i} &  =\left(  \mathcal{A}Q^{(r)}+\mathcal{B}%
N^{(r)}\right)  \sigma_{ij}^{(r)}\varepsilon^{j},\nonumber\\
Q^{(r)}Q^{(r)} &  =N^{(r)}N^{(r)}=1,\text{ \ \ \ }\mathcal{A}^{2}%
+\mathcal{B}^{2}=1,\text{ \ \ \ \ \ }Q^{(r)}N^{(r)}=0,\label{newproj}%
\end{align}
where all quantities appearing in the projection condition are in general
field dependent. If $\mathcal{A}=1,$ one has the projection condition of
\cite{domaink}. For these cases, as mentioned in the previous section, it was
found that one must satisfy $\partial_{x}Q^{(r)}=0$ in order to obtain
Ricci-flat BPS domain wall solutions.

The metric of our curved domain wall can be put in the form
\begin{equation}
ds^{2}=e^{2U(z)}g_{\mu\nu}(x)dx^{\mu}dx^{\upsilon}+dz^{2},\label{ansatz}%
\end{equation}
and all the dynamical scalar fields of the theory are assumed to depend only
on the fifth coordinate $z$. The non-vanishing spin connections for our metric
are given by
\begin{align}
\Omega_{\mu ab}(x,z) &  =\omega_{\mu ab}(x),\nonumber\\
\Omega_{\mu a5}(x,z) &  =U^{\prime}e^{U}e_{a\mu}(x).
\end{align}
From the vanishing of the $\mu$-component of the gravitini supersymmetry
transformation we obtain
\begin{equation}
\delta\psi_{\mu i}=D_{\mu}\varepsilon_{i}+\frac{1}{2}e^{U}\gamma_{\mu}\left(
(\mathcal{A}U^{\prime}-gW)Q^{\left(  r\right)  }+\mathcal{B}U^{\prime}%
N^{(r)}\right)  \sigma_{ij}^{\left(  r\right)  }\varepsilon^{j},\label{jihad}%
\end{equation}
where we use the projection condition (\ref{newproj}) as well as
\begin{equation}
\Gamma_{\mu}=e^{U}\gamma_{\mu},\ \ \Gamma_{z}=\gamma_{5},\ \ \ \ \ \ D_{\mu
}=\partial_{\mu}+\frac{1}{4}\omega_{\mu}^{ab}\gamma_{ab}.
\end{equation}
The vanishing of the gaugini supersymmetry transformation results in the
following equations representing the supersymmetric flow of the
vectormultiplets scalars \cite{lust}
\begin{align}
\mathcal{A}\phi^{\prime x} &  =-3gg^{xy}\partial_{y}W,\label{d}\\
\mathcal{B}N^{\left(  r\right)  }\phi^{\prime x} &  =-3gg^{xy}W\partial
_{y}Q^{(r)}.\label{dd}%
\end{align}
Notice that these equations \ generalize the first order differential equation
given in \cite{karch} to the cases of many vectormultiplets.

From the vanishing of the hyperini supersymmetry transformation, we obtain the
supersymmetric flow equation of the hyper scalars. This is given by
\begin{equation}
\left(  \mathcal{A}g_{XY}+2\mathcal{B}\epsilon^{(r)(s)(t)}Q^{(r)}N^{(s)}%
R_{XY}^{(t)}\right)  q^{\prime Y}=\mathcal{A}G_{XY}q^{\prime Y}=-3g\partial
_{X}W.\label{ddd}%
\end{equation}
If one requires that the worldvolume Ricci-tensor satisfy
\begin{equation}
R_{\mu\nu}^{(4)}=-12c^{2}g_{\mu\nu},
\end{equation}
then integrability coming from the vanishing of (\ref{jihad}) will imply
\cite{lust}
\begin{equation}
(\mathcal{A}U^{\prime}-gW)^{2}+\mathcal{B}^{2}U^{\prime2}=4c^{2}%
e^{-2U}.\label{nice}%
\end{equation}
We now turn to the analysis of the equations of motion. The Einstein equations
of motion give for the worldvolume Ricci-tensor
\begin{equation}
R_{\mu\nu}^{(4)}=g_{\mu\nu}e^{2U}(4U^{\prime2}+U^{\prime\prime}+\frac{2}%
{3}\mathcal{V)}.\label{general ricci}%
\end{equation}
Also one finds the following expression for the $zz$-component of the five
dimensional Ricci-tensor
\begin{equation}
R_{zz}^{(5)}=g_{\Lambda\Sigma}\partial_{z}\phi^{\Lambda}\partial_{z}%
\phi^{\Sigma}+\frac{2}{3}\mathcal{V}=-4(U^{\prime\prime}+U^{\prime
2}).\label{scorpion}%
\end{equation}
Using (\ref{d}) and (\ref{dd}), the scalar potential of the theory now takes
the form \cite{lust}
\begin{align}
\mathcal{V} &  =g^{2}(-6W^{2}+\frac{9}{2}g^{\Lambda\Sigma}\partial_{\Lambda
}W\partial_{\Sigma}W+\frac{9}{2}W^{2}g^{xy}\partial_{x}Q^{(r)}\partial
_{y}Q^{(r)}\mathcal{)}\nonumber\\
&  =g^{2}(-6W^{2}+\frac{9}{2}g^{XY}\partial_{X}W\partial_{Y}W+\frac
{9}{2\mathcal{A}^{2}}g^{xy}\partial_{x}W\partial_{y}W\mathcal{)}%
.\label{spotential}%
\end{align}
From (\ref{general ricci}) and (\ref{scorpion}) together with the
supersymmetric flow equations, one finally gets for the worldvolume
Ricci-tensor the following expression
\begin{equation}
R_{\mu\nu}^{(4)}=3g_{\mu\nu}e^{2U}\left(  U^{\prime2}-g^{2}W^{2}-\frac{3}%
{4}g^{2}\partial_{X}W\partial_{Y}W(\frac{1}{\mathcal{A}^{2}}G^{YV}G^{XU}%
g_{UV}-g^{XY})\right)  .\label{qmw}%
\end{equation}
The equation of motion for the scalar fields derived from the action
(\ref{smm}) reads
\begin{equation}
\frac{\partial}{\partial\phi^{\Lambda}}\mathcal{V}+\frac{1}{2}\partial
_{\Lambda}g_{\Gamma\Sigma}\phi^{\prime\Gamma}\phi^{\prime\Sigma}%
=e^{-4U}(e^{4U}g_{\Lambda\Sigma}\phi^{\prime\Sigma})^{\prime}.\label{eqmm}%
\end{equation}
Using the expression for $\mathcal{V}$ as given in (\ref{spotential}), one
then obtains, respectively, for the vectormultiplets and hypermultiplets
scalars, the following equations
\begin{align}
&  12g\left(  \frac{U^{\prime}}{\mathcal{A}}-gW\right)  \partial_{z}%
W+\frac{9g^{2}}{\mathcal{A}^{3}}g^{xy}\partial_{y}W\mathcal{(}\text{ }%
\partial_{x}\mathcal{A}\partial_{z}W-\partial_{x}W\partial_{z}\mathcal{A}%
)\nonumber\\
&  +9g^{2}\partial_{Y}W\partial_{X}\partial_{z}W(g^{XY}-\frac{G^{XY}%
}{\mathcal{A}^{2}})+\frac{9g^{2}}{\mathcal{A}^{3}}G^{XY}\text{ }\partial
_{X}\mathcal{A}\partial_{Y}W\partial_{z}W=0,\label{fa}%
\end{align}%
\begin{align}
&  12g(\frac{U^{\prime}}{\mathcal{A}}g_{ZX}G^{XY}-gW\delta_{Z}^{Y}%
)\partial_{Y}W+\frac{9g^{2}}{2}\partial_{Z}g_{XY}\partial_{U}W\partial
_{V}W(\frac{1}{\mathcal{A}^{2}}G^{XU}G^{YV}-g^{XU}g^{YV})\nonumber\\
&  +9g^{2}\partial_{U}\partial_{Y}W\partial_{V}W(\delta_{Z}^{U}g^{YV}-\frac
{1}{\mathcal{A}^{2}}g_{ZX}G^{XY}G^{UV})\nonumber\\
&  +\frac{9g^{2}}{\mathcal{A}^{2}}\partial_{x}\partial_{Y}W\partial_{y}%
Wg^{xy}(\delta_{Z}^{Y}-g_{ZX}G^{XY})\mathcal{+}\frac{9g^{2}}{\mathcal{A}^{3}%
}g_{ZX}G^{XY}\partial_{Y}WG^{UV}\partial_{U}\mathcal{A}\partial_{V}%
W\nonumber\\
&  -\frac{9g^{2}}{\mathcal{A}^{3}}g^{xy}\partial_{Z}\mathcal{A}\partial
_{x}W\partial_{y}W+\frac{3g}{\mathcal{A}}(g_{ZX}G^{XY})^{\prime}\partial
_{Y}W\mathcal{+}\frac{9g^{2}}{\mathcal{A}^{3}}g_{ZX}G^{XY}\partial_{Y}%
Wg^{xy}\partial_{x}\mathcal{A}\partial_{y}W\mathcal{\ }\nonumber\\
&  =0.\label{faa}%
\end{align}
In the following, we will demonstrate that some of the conditions one must
impose in order to obtain Einstein domain walls are
\begin{equation}
\text{\ \ }\partial_{X}W=0,\text{ \ \ \ }\partial_{X}\mathcal{A=}0.\label{k}%
\end{equation}
\ \ For such conditions we obtain, from the Ricci-scalar equation (\ref{qmw})
as well as integrability condition\ (\ref{nice}), the equations
\begin{align}
(U^{\prime2}-g^{2}W^{2}) &  =-4c^{2}e^{-2U},\label{aids}\\
U^{\prime2}+g^{2}W^{2}-2g\mathcal{A}WU^{\prime} &  =4c^{2}e^{-2U},\label{ai}%
\end{align}
which, for non-trivial warp factor, imply that
\begin{align}
U^{\prime} &  =\mathcal{A}gW,\nonumber\\
\mathcal{B}^{2}g^{2}W^{2}e^{2U} &  =4c^{2}.\label{jk}%
\end{align}
These equations agree with the modified equation for the warp factor as
suggested in \cite{karch}. Going back to the equations of the scalar fields
and using (\ref{jk}) and (\ref{k}), it can be easily seen that (\ref{fa}) is
satisfied provided
\begin{equation}
\text{\ \ \ \ \ }\mathcal{(}\partial_{x}\mathcal{A}\partial_{z}W-\partial
_{x}W\partial_{z}\mathcal{A})=0.\label{conn}%
\end{equation}
From the hyper scalars equation of motion (\ref{faa}) we get the condition
\begin{equation}
\partial_{x}\partial_{Y}W\partial_{y}Wg^{xy}(\delta_{Z}^{Y}-g_{ZX}G^{XY})=0,
\end{equation}
and therefore, if $g_{ZX}G^{XY}\neq\delta_{Z}^{Y},$ one must impose the
condition $\partial_{x}\partial_{Y}W=0.$ For gauged supergravity theories with
one vectormultiplet, the condition (\ref{conn}) is automatically satisfied.
Moreover, using (\ref{k}), the scalar potential of the theory becomes
\begin{equation}
\mathcal{V}=g^{2}(-6W^{2}+\frac{9}{2\mathcal{A}^{2}}g^{xy}\partial
_{x}W\partial_{y}W\mathcal{)},
\end{equation}
which agrees with the form of the scalar potential given in \cite{karch}. Note
that the equations of motion only require that the worldvolume be given by an
Einstein metric which may or may not admit any supersymmetry.

We now return to the analysis of the projection condition (\ref{newproj}). The
vanishing of the fifth component of the gravitini supersymmetry variation
together with (\ref{jihad}) imply more integrability conditions given by
\begin{equation}
e^{U}(\mathcal{A}U^{\prime}-gW)Q^{(r)}+e^{U}\mathcal{B}U^{\prime}%
N^{(r)}=2c^{\left(  r\right)  },\label{moreinteg}%
\end{equation}
where $c^{\left(  r\right)  }=c_{0}^{\left(  s\right)  }O^{sr}$,
$c_{0}^{\left(  s\right)  }$ are constants and $O^{sr}$ is an orthogonal
matrix given by%
\begin{align*}
O^{sr}  & =\cos2\alpha\delta^{sr}+2\sin^{2}\alpha\frac{\alpha^{s}\alpha^{r}%
}{\alpha^{2}}-\epsilon^{srt}\sin2\alpha\frac{\alpha^{t}}{\alpha},\\
\alpha^{r}  & =ce^{-U}\epsilon^{rst}Q^{s}N^{t}-\frac{i}{2}q^{\prime X}%
\omega_{Xi}^{\quad j}\left(  \sigma^{r}\right)  _{j}^{i},\\
\alpha^{2}  & =\alpha^{r}\alpha^{r}.
\end{align*}
Notice that $c^{\left(  r\right)  }c^{\left(  r\right)  }=c_{0}^{\left(
r\right)  }c_{0}^{\left(  r\right)  }=c^{2}$ and $O^{sr}O^{tr}=\delta^{st}.$

From the `supersymmetric' flow equations (\ref{d}) and (\ref{dd}) one obtains,
for non-trivial scalars, the following condition%

\begin{equation}
\mathcal{B}N^{(r)}\partial_{x}W=\mathcal{A}W\partial_{x}Q^{(r)}.\label{ka}%
\end{equation}
Using (\ref{jk}), (\ref{moreinteg}) we obtain
\begin{equation}
\pm c\left(  \mathcal{A}N^{(r)}-\mathcal{B}Q^{(r)}\right)  =c^{(r)}%
\end{equation}
and upon using (\ref{newproj}), one finds that%

\begin{align}
\mp c\mathcal{B} &  =c^{(r)}Q^{(r)},\nonumber\\
\pm c\mathcal{A} &  =c^{(r)}N^{(r)}.\label{kaa}%
\end{align}
Multiplying (\ref{ka}) by $c^{(r)}$ (and summing over $r)$ , then using
(\ref{kaa}) one can finally derive the following condition:%

\begin{equation}
\partial_{x}(\mathcal{B}W)=-\frac{c_{0}^{\left(  s\right)  }}{c}WQ^{\left(
r\right)  }\partial_{x}O^{sr}.\label{dis}%
\end{equation}

In what follows we give examples of possible Einstein domain wall solutions.
From the condition (\ref{conn}), it can be seen that solutions may exist if
one takes $\mathcal{A}=\mathcal{A}(W)$. Using (\ref{jk}) and (\ref{aids}) we
obtain
\begin{align}
4c^{2}e^{-2U}  & =G(W),\label{kod}\\
g^{2}W^{2}(1-\mathcal{A}^{2})  & =g^{2}W^{2}\mathcal{B}^{2}%
=G(W).\label{kodtwo}%
\end{align}
where $G$ is some function of $W.$ Differentiating equation (\ref{kod}) one
arrives at the following equation
\begin{equation}
\frac{2}{3}\mathcal{A}^{2}WG=(g^{xy}\partial_{x}W\partial_{y}W)\frac{dG}{dW}%
\end{equation}
which is only consistent if one allows for the very restrictive condition
$\ g^{xy}\partial_{x}W\partial_{y}W=f(W)$ in which case one has
\begin{equation}
\frac{1}{G}\frac{dG}{dW}=\frac{2}{3}\frac{\mathcal{A}^{2}W}{f}=\frac{2W}%
{3f}-\frac{2G}{3g^{2}fW},\label{fff}%
\end{equation}
which in reality is nothing but a differential equation for $\mathcal{A}.$
Differentiating (\ref{kodtwo}) and using (\ref{fff}) one arrives at the
following differential equation for $\mathcal{A},$
\begin{equation}
\frac{\mathcal{A}}{1-\mathcal{A}^{2}}\frac{d\mathcal{A}}{dW}=-\frac{1}{3}%
\frac{\mathcal{A}^{2}W}{f}+\frac{1}{W},\text{ \ \ }%
\end{equation}
This can be solved by
\begin{equation}
\mathcal{B}^{2}=\frac{3}{2}\frac{e^{2\int(\frac{W}{3f(W)}-\frac{1}{W})dW}%
}{\int dW\frac{W}{f}e^{2\int(\frac{W}{3f(W)}-\frac{1}{W})dW}}=1-\mathcal{A}%
^{2}.
\end{equation}
Equivalently we can solve directly for the warp factor. From equation $\left(
\ref{kod}\right)  $, it is clear that the knowledge of $G(W)$ fixes $e^{-2U}.$
The differential equation for $G$ given by (\ref{fff}) can be integrated and
one gets the following solution
\begin{equation}
e^{2U}=4c^{2}\frac{1}{G}=\frac{8c^{2}}{3g^{2}}e^{-\int\frac{2W}{3f(W)}dW}%
\int\frac{1}{Wf}e^{\int\frac{2W}{3f(W)}dW}dW.\label{fadia}%
\end{equation}
Obviously, the dependence of $U$ on the coordinate $z$ is determined by
$W=W(z).$ Recalling that
\begin{equation}
W^{\prime}=\partial_{x}W\phi^{\prime x}=-\frac{3g}{\mathcal{A}}g^{xy}%
\partial_{x}W\partial_{y}W=-\frac{3g}{\mathcal{A}}f,
\end{equation}
we then obtain the following relation
\begin{equation}
\int\frac{\mathcal{A}dW}{f(W)}=-3g(z-z_{0}).
\end{equation}

\section{Discussion}

In this paper we studied the possibility of constructing domain wall solutions
with Einstein worldvolumes (with a negative cosmological constant) for the
theories of five dimensional $N=2$ gauged \ supergravity of \cite{ceresole} in
the absence of tensormultiplets. The first order differential equations
representing the so-called supersymmetric flow of the scalars are derived (see
also \cite{lust}). These equations are obtained by assuming a certain
projection condition on the Killing spinors of the BPS solution and solving
for the vanishing of supersymmetry transformation of the fermionic fields in a
bosonic background. The supersymmetric flow equations of the scalars together
with Einstein and the scalar fields equations of motion are analyzed and
conditions under which solutions with a cosmological constant on the
worldvolume may exist are determined. It turns out that these conditions can
be made compatible with integrability conditions coming from the Killing
spinor equations provided certain conditions are satisfied.

The main result of this paper is the derivation of the constraints that must
be imposed on the supergravity model in order to have domain walls with
Einstein worldvolumes. The possible solutions considered generalize those
considered in \cite{karch} to an arbitrary number of vector and
hypermultiplets. It remains to be seen whether one can construct explicitly
$N=2$ \ gauged supergravity theories satisfying the constraints derived in
this paper which are necessary for the existence of curved Einstein domain
walls. Moreover, it would be interesting to generalize our results to four
dimensional theories and investigate general domain wall solutions with
non-trivial gauge fields on the worldvolume \cite{branes}. We hope to report
on these issues in a future publication. \bigskip

\textbf{Acknowledgments}. We would like to thank Dieter L\"{u}st for
correspondence. W. A. S thanks J. M. Figueroa-O'Farrill for a useful discussion.

\end{document}